\documentclass[10pt]{article}
\pdfoutput=1
\usepackage{epic,graphpap,graphics,float,tabularx}
\usepackage{paralist,longtable,fancyhdr,fancyvrb,bbm,natbib}
\usepackage{amsmath}
\usepackage{amstext}
\usepackage{amsbsy}
\usepackage{amsthm}
\usepackage{amsfonts}
\usepackage{amssymb}
\usepackage{amscd}
\usepackage{eucal}
\usepackage{comment}

\theoremstyle{plain}
\newtheorem{thm}{Theorem}[section]
\newtheorem{lem}[thm]{Lemma}

\newtheorem{prop}[thm]{Proposition}
\theoremstyle{definition}
\newtheorem{defn}[thm]{Definition}
\newtheorem{exa}[thm]{Example}

\numberwithin{equation}{section}
\def\as{\quad\text{{\rm a.s.}}}
\def\sumi{\sum_{i=1}^n}
\def\sumj{\sum_{j=1}^n}
\def\sumk{\sum_{k=1}^n}
\def\a{\alpha}
\def\m{\mu}

\def\t{\tau}

\def\g{\gamma}

\def\s{\sigma}
\def\r{\rho}

\def\XX{{\mathcal X}}

\def\1{{\mathbbm 1}}
\def\E{{\mathbb E}}
\def\N{{\mathbb N}}

\def\R{{\mathbb R}}
\def\eqdef{\triangleq}
\def\bg{\boldsymbol g}

\def\bs{\boldsymbol \sigma}
\def\bl{\boldsymbol \lambda}
\def\l{\lambda}

\def\sumi{\sum_{i=1}^n}

\def\half{\frac{1}{2}}
\def\brac#1{\langle #1\rangle}
\def\bbrac#1{\big\langle #1 \big\rangle}
\def\lt#1{\Lambda_{#1}}
\def\ito{It\^o}
\def\limT#1{\lim_{T\to\infty}\frac{#1}{T}}
\def\limt#1{\lim_{t\to\infty}\frac{#1}{t}}

\def\intT{\int_0^T}

\begin{document}

\centerline{\bf \LARGE{Zipf's law for Atlas models}}

\vskip 10pt

\centerline{\large{Ricardo T. Fernholz\footnote[1]{Claremont McKenna College, 500 E. Ninth St., Claremont, CA 91711, rfernholz@cmc.edu.} \hskip 25pt Robert Fernholz\footnote[2]{Intech Investment Management, LLC, One Palmer Square, Princeton, NJ 08542, bob@bobfernholz.com.}}}

\vskip 20pt

\centerline{\large{\today}}

\vskip 100pt

\begin{abstract}
A set of data with positive values follows a {\em Pareto distribution} if the log-log plot of value versus rank is approximately a straight line.  A Pareto distribution satisfies {\em Zipf's law} if the log-log plot has a slope of~$-1$. Since many types of ranked data follow Zipf's law, it is considered a form of universality. We propose a mathematical explanation for this phenomenon based on Atlas models and first-order models, systems of strictly positive continuous semimartingales with parameters that depend only on rank. We show that the stationary distribution of an Atlas model will follow Zipf's law if and only if two natural conditions, conservation and completeness, are satisfied.  Since Atlas models and first-order models can be constructed to approximate systems of time-dependent rank-based data, our results can explain the universality of Zipf's law for such systems. However,  ranked data generated by other means may follow non-Zipfian Pareto distributions. Hence, our results explain why Zipf's law holds for word frequency, firm size, household wealth, and city size, while it does not hold for earthquake magnitude, cumulative book sales, the intensity of solar flares, and the intensity of wars, all of which follow non-Zipfian Pareto distributions.
\end{abstract}

\vfill

{\em MSC2010 subject classifications:} 60H30, 91B70, 91G70, 91D20, 93E03


\pagebreak

\section{Introduction} \label{intro}

A set of empirical data with positive values follows a {\em Pareto distribution} if the log-log plot of the values versus rank is approximately a straight line. Pareto distributions are ubiquitous in the social and natural sciences, appearing in a wide range of fields from geology to economics \citep{Simon:1955,Bak:1996,Newman:2005}. A Pareto distribution satisfies {\em Zipf's law} if the log-log plot has a slope of~$-1$, following \citet{Zipf:1935}, who noticed that the frequency of written words in English follows such a distribution. We shall refer to these distributions as {\em Zipfian}. Zipf's law is considered a form of universality, since Zipfian distributions occur almost as frequently as Pareto distributions. Nevertheless, according to  \citet{Tao:2012}, ``mathematicians do not have a fully satisfactory and convincing explanation for how the law comes about and why it is universal.'' 

We propose a mathematical explanation of Zipf's law based on {\em Atlas models} and {\em first-order models}, systems of strictly positive continuous semimartingales with parameters that depend only on rank. Atlas and first-order models were introduced by \citet{F:2002} to model the distribution of capital in stock markets, and a mathematical development of these models can be found in \citet{BFK:2005}, \citet{FK:2009}, and \citet{IPBKF:2011}. Atlas and first-order models can be constructed to approximate empirical systems of time-dependent rank-based data that exhibit some form of stability, and while the stationary distributions of Atlas models are Pareto, first-order models can be constructed to have any stationary distribution \citep{F:2002}. 

Many empirical systems of time-dependent rank-based data generate distributions with log-log plots that are not actually straight lines but rather are concave curves with a tangent of slope $-1$ at some point along the curve.  We shall refer to these more general distributions as {\em quasi-Zipfian,} and we shall use first-order models to approximate the systems that generate them. 

The class of empirical systems for which Zipf's law, or its quasi-Zipfian counterpart, is likely to hold comprises large time-dependent systems for which the number of members can vary over time. Frequency of written words in a language, population of cities, and capitalization of U.S. companies all fall into this class. These systems frequently satisfy two natural conditions, {\em conservation} and {\em completeness}. Conservation is like conservation of mass in a physical system, and arises, for example, in measuring the frequency of written words. Since it is impossible to count all the written words in a language, a given number of words must be sampled, and conservation is the result of maintaining a constant sample size over time. Hence, conservation is a natural condition that can be expected to hold for many time-dependent rank-based systems of empirical data.

The second condition, completeness, is related to the replacement of members at the bottom of a rank-based empirical system. In a large rank-based system of time-dependent data those members in the lowest ranks will frequently be replaced by new members from outside the system, and completeness ensures that the effect of this replacement is minimal if the system includes enough ranks. As an example, in Section~\ref{disc} we show that the distribution of capital in the U.S. stock market follows a complete quasi-Zipfian distribution. However, if this distribution is cut off after the top 100 stocks, the resulting incomplete system is no longer quasi-Zipfian. While it is certainly possible to construct incomplete systems, like the top 100 stocks, most such systems seem to be truncated versions of larger complete systems. Accordingly, conservation and completeness are broadly universal properties of large systems of time-dependent rank-based empirical data.

Mathematically, we show that under the assumptions of conservation and completeness, the stationary distribution of an Atlas model will satisfy Zipf's law. However, most time-dependent rank-based systems do not quite satisfy Zipf's law, and also do not quite satisfy the requirements for Atlas models, so in practice we usually must employ more general first-order models. We refer to these more general models as {\em quasi-Atlas models,} and we show that under conservation and completeness these models will result in quasi-Zipfian distributions as long as the top-ranked process represents less than half the total mass of the system. Quasi-Atlas models can be used to approximate many large rank-based systems, and since conservation and completeness are common characteristics of such systems, this offers an explanation for the universality of quasi-Zipfian distributions in the natural and social sciences.

The dichotomy between the class of Zipfian and quasi-Zipfian distributions versus the class of non-Zipfian Pareto distributions is of interest to us here. We find that Zipfian and quasi-Zipfian distributions are usually generated by systems of time-dependent rank-based data, and it is this class of systems that we can approximate by Atlas models or first-order models. In contrast, data that follow non-Zipfian Pareto distributions are usually generated by other means, often of a cumulative nature. Examples of time-dependent rank-based systems that generate Zipfian or quasi-Zipfian distributions include the market capitalization of companies \citep{Simon/Bonini:1958,F:2002}, the population of cities \citep{Gabaix:1999}, the employees of firms \citep{Axtell:2001}, the income and wealth of households \citep{Atkinson/Piketty/Saez:2011,Piketty:2017}, and the assets of banks \citep{Fernholz/Koch:2017}. From the comprehensive survey of \citet{Newman:2005} we find an assortment of non-Zipfian Pareto distributions:  the magnitude of earthquakes, citations of scientific papers, copies of books sold, the diameter of moon craters, the intensity of solar flares, and the intensity of wars, all of which are cumulative systems. Consider, for example,  the magnitude of earthquakes: each new earthquake adds a new observation to the data, but once recorded, these observations do not change over time. Such cumulative systems may generate Pareto distributions, but we have no reason to believe that these distributions will be Zipfian.

The mathematical theory of Atlas models and first-order models developed in \citet{BFK:2005} and \citet{IPBKF:2011} is based on a number of earlier results. The existence and uniqueness for solutions of these systems comes from \citet{Bass/Pardoux:1987} and \citet{Stroock/Varadhan:2006}. The behavior of the ``gap processes,'' the differences between adjacent rank processes, is based on the work of \citet{Harrison/Reiman:1981}, \citet{Harrison/Williams:1987a,Harrison/Williams:1987b}, and \citet{Williams:1987}. The long-term behavior of Atlas models and first-order models, including the existence of a stationary distribution and a strong law of large numbers, can be found in \citet{Khasminskii:1960,Khasminskii:1980}.

 The theory of rank-based systems of continuous semimartingales has been extended in several directions, e.g., infinite Atlas systems \citep{Pal/Pitman:2008,Chatterjee/Pal:2013,Bruggeman:2016}; behavior at triple points \citep{Banner/Ghomrasni:2008}; existence and nonexistence of triple points \citep*{IK:2010,IchibaKaratzasShkolnikov:2013,Sarantsev:2015}; convergence to equilibrium \citep*{IchibaPalShkolnikov-2013,Dembo/Jara/Olla:2017,Dembo/Tsai:2017}; behavior of degenerate systems \citep*{FIK:2013b,FIKP:2013}; large deviations \citep{DSVZ:2016}; and second-order stock-market models \citep*{FIK:2013a}.

In the next sections we first review the properties of Atlas models and first-order models and then characterize Zipfian and quasi-Zipfian systems using these models. We apply our results to the capitalization of U.S.\ companies, with an analysis of the corresponding quasi-Zipfian distribution curve. We also discuss a number of other time-dependent systems as well as other approaches that have been used to characterize these systems. Proofs of all propositions are in the appendix.

\section{Atlas models and quasi-Atlas models} \label{atlas}

We use systems of strictly positive continuous semimartingales $\{X_1,\ldots, X_n\}$, with $n>1$, to approximate systems of time-dependent data. For such a system we define the {\em rank function} to be the random permutation $r_t\in\Sigma_n$ such that $r_t(i)<r_t(j)$ if $X_i(t)>X_j(t)$ or if $X_i(t)=X_j(t)$ and $i<j$. Here $\Sigma_n$ is the symmetric group on $n$ elements. The {\em rank processes} $\{ X_{(1)}\ge\cdots\ge X_{(n)} \}$ are defined by $X_{(r_t(i))}(t)=X_i(t)$.

For a continuous semimartingale $X$, we can define the {\em semimartingale local time at the origin} $\lt{}$ by the Tanaka-Meyer formula
\[
\lt{}(t)\eqdef \half\Big(|X(t)|-|X(0)|-\int_0^t\text{sgn}(X(s))\,dX(s)\Big),
\]
for $t\ge0$, where sgn$(x)=2\,\1_{\{x>0\}}-1$, for $x\in\R$ (see (7.7)--(7.9) of \citet{Karatzas/Shreve:1991}, page 220). The local time $\lt{}$ measures the amount of time that $X$ spends near $0^+$. The mapping $t \mapsto \lt{}(t)$ is continuous and non-decreasing and induces the random measure  $d\lt{}$ with support contained in the set $\{t\ge0:X(t)=0\}$ (\citet{Karatzas/Shreve:1991}, Theorem 7.1(ii), page 218).

We have assumed that the semimartingales $X_i(t)$ are strictly positive, so we can consider the logarithmic processes $\log X_1,\ldots,\log X_n$. For $1\le k< \ell \le n$, let $\lt{k,\ell}^X$ denote  the local time at the origin for $\log X_{(k)}-\log X_{(\ell)}$, with $\lt{0,1}^X=\lt{n,n+1}^X\equiv 0$. The processes $\log X_1,\ldots,\log X_n$ have a {\em triple point} at time $t>0$ if there exist $j<k<\ell$ such that $\log X_j(t)=\log X_k(t)=\log X_{\ell}(t)$. Multidimensional Brownian motion almost surely has no triple points (see \citet{Karatzas/Shreve:1991}, Proposition~3.22, page 161), but some of the systems we consider satisfy only the weaker condition that the processes $\log X_1,\ldots,\log X_n$ {\em accumulate no local time at triple points,} by which we mean that  for all $\ell\ge k+2$, we have $\lt{k,\ell}\equiv 0$, a.s. If the $\log X_i$ accumulate no local time at triple points, then Theorem 2.5 of \citet{Banner/Ghomrasni:2008} shows that the rank processes $\log X_{(k)}$ satisfy
\begin{equation}\label{2.1}
d\log X_{(k)}(t)=\sumi\1_{\{r_t(i)=k\}}\,d\log X_i(t)+\half d\lt{k,k+1}^X(t)-\half d\lt{k-1,k}^X(t),\as,
\end{equation}
for $k = 1, \ldots, n$.  

Let us define the processes
\begin{equation*}
X_{[k]} \eqdef X_{(1)}+\cdots+X_{(k)},
\end{equation*}
for $k=1,\ldots,n$. The following lemma shows that the local time process $\lt{k,k+1}^X$ measures the flow into and out of $X_{[k]}$.

\begin{lem} \label{L2.1}
Let $X_1, \ldots, X_n$ be strictly positive continuous semimartingales that satisfy \eqref{2.1}. Then
\begin{equation}\label{2.2}
\half X_{(k)}(t)d\lt{k,k+1}^X(t)= dX_{[k]}(t)- \sum_{i=1}^n \1_{\{r_t(i)\le k\}}dX_i(t),\as,
\end{equation}
for $k=1,\ldots,n$.
\end{lem}


The local time process $\lt{k,k+1}^X$ compensates for turnover into and out of the top $k$ ranks. Over time, some of the higher-ranked processes will decrease and exit from the top ranks, while some of the lower-ranked processes will increase and enter those top ranks. Equation \eqref{2.2}  measures the replacement of the top $k$ ranks of the system by the lower ranks.

We are interested in systems that show stability by rank, at least asymptotically. Since we must apply our definition of stability to systems of empirical data as well as to continuous semimartingales, we use asymptotic time averages rather than expectations for our definitions. We shall show below that for the systems of continuous semimartingales we consider, a law of large numbers implies that the asymptotic time averages are equal to the corresponding expectations.

\begin{defn} \label{D2.2}
\citep{F:2002} Let $\{X_1, \ldots, X_n\}$ be a system of  strictly positive continuous semimartingales that satisfy \eqref{2.1}. Then this system is {\em asymptotically stable} if there exist positive constants $\l_{k,k+1}$ and $\s^2_{k,k+1}$,  $k=1,\ldots,n-1$, such that
\begin{enumerate}
\item $\displaystyle
       \limt{1}\big( \log X_{(1)}(t)-\log X_{(n)}(t)\big)=0,\as$ ({\em coherence});
\item $\displaystyle
       \limt{1}\lt{k,k+1}^X(t) = \l_{k,k+1},\as, \text{ for }  k=1,\ldots,n-1$;
\item $\displaystyle
      \limt{1}\bbrac{\log X_{(k)}-\log X_{(k+1)}}_t =
        \s^2_{k,k+1},\as, \text{ for } k=1,\ldots,n-1$;
\end{enumerate}
where $\brac{\,\cdot\,}$ represents quadratic variation. 
\end{defn}\vspace{5pt}

The simplest system we consider is an {\em Atlas model}, a system of strictly positive continuous semimartingales $\{X_1,\ldots,X_n\}$ defined by
\begin{equation}\label{2.3}
d\log X_i(t) = -g\,dt+ng\1_{\{r_t(i)=n \}}dt+\s\,dW_i(t),
\end{equation}
for $i=1,\ldots,n$, where $g>0$ and $\s>0$ are constants and $(W_1, \ldots, W_n)$ is a Brownian motion  (see Example~5.3.3 of \citet{F:2002}, page~103).  Atlas models are asymptotically stable with parameters
\begin{equation} \label{2.4}
\l_{k,k+1} = 2kg \quad\text{ and }\quad \s^2_{k,k+1} =2 \s^2, 
\end{equation}
for $k = 1, \ldots, n-1$ (see Proposition~2 of \citet{IPBKF:2011}). 

A modest generalization of the Atlas model is the first-order model, introduced in Section~5.5 of \citet{F:2002}. A {\em first-order model} is a system of  strictly positive continuous semimartingales $\{X_1,\ldots,X_n\}$ with
\begin{equation}\label{2.5}
d\log X_i(t) = g_{r_t(i)}\,dt+G_n\1_{\{r_t(i)=n\}}dt+\s_{r_t(i)}\,dW_i(t),
\end{equation}
for $i=1,\ldots,n$, where $\s^2_1,\ldots,\s^2_n$ are positive constants;  $g_1,\ldots,g_n$ are constants that satisfy
\begin{equation}\label{2.51}
 g_1+\cdots+g_n \le 0 \quad\text{ and }\quad g_1+\cdots+g_k < 0 \text{ for } k < n;
\end{equation}
$G_n=-(g_1 + \cdots + g_n)$; and $(W_1, \ldots, W_n)$ is a Brownian motion (see (1.1)--(1.6) of \citet{BFK:2005}). First-order models are asymptotically stable with parameters
\begin{equation} \label{2.6}
\l_{k,k+1} = -2\big(g_1+\cdots+g_k\big), \as,
\end{equation}
and
\begin{equation}\label{2.7}
\s^2_{k,k+1} = \s_k^2+\s^2_{k+1}, \as,
\end{equation}
for $k = 1, \ldots, n-1$ (see Proposition~2 of \citet{IPBKF:2011}). Here we use a simple form of first-order model in which the drift parameters $g_k$ are constant and the variance parameters $\s^2_k$ grow linearly with rank. Accordingly, we define a {\em quasi-Atlas model} to be a first-order model determined by three parameters $g>0$ and $\s^2_2\ge\s^2_1>0$, such that
\begin{align}
g_k&=-g,\label{2.49}\\
\s^2_k&=\s^2_1+(k-1)(\s^2_2-\s^2_1),\label{2.50}
\end{align}
for $k=1,\ldots,n$. Hence, we see that Atlas models are a subclass of quasi-Atlas models, which in turn are a subclass of first-order models. 

By Proposition~2.3 of \citet{BFK:2005}, each of the processes $X_i$ in a first-order model asymptotically spends equal time in each rank. Due to this ergodicity, the parameters $ng$  in \eqref{2.3} and $G_n$ in \eqref{2.5} cause the asymptotic growth rate to be zero for each of the processes $\log X_i$, for $i=1,\ldots,n$. Equations \eqref{2.3} and \eqref{2.5} can be generalized by the addition of a term $\g\,dt$ on the right-hand side, where the constant $\g$ represents the common logarithmic growth rate of the system, but in our setting it is convenient to make the simplifying assumption that $\g=0$ (see, e.g., \citet{BFK:2005}, equations (1.1) and  (1.6)). The condition \eqref{2.51}, along with $G_n=-(g_1 + \cdots + g_n)$, stabilizes the system and prevents it from separating into smaller subsystems over time. A discussion of this stabilizing effect can be found in the Remark following Theorem~8 of \citet{Pal/Pitman:2008}.

We see from \eqref{2.6} and \eqref{2.7} that for a first-order model the parameters $\l_{k,k+1}$ and $\s^2_{k,k+1}$ depend only on ranks 1 through $k+1$ and not on the number $n$ of processes in the model. On a more intuitive level, the parameter $G_n$ is defined so that whatever the size $n$ of the model, the ``upward force''  $g_{k+1}+\cdots+g_n+G_n>0$ from below  adjusts  to  counteract the ``downward force'' $g_1+\cdots+g_k<0$  from above,
with
\[
g_{k+1}+\cdots+g_n+G_n=-(g_1+\cdots+g_k).
\]
The local time $\lt{k,k+1}$ between ranks $k$ and $k+1$ is determined by these upward and downward forces since they push these two ranks  together, and  the value of $\l_{k,k+1}$ depends on this local time. 

Lemma~1 of \citet{IPBKF:2011}  shows that the processes $\log X_1,\ldots,\log X_n$ in a first-order model accumulate no local time at triple points. It is also known that a first-order model for which  $k\mapsto\s^2_k$ is {\em concave,} i.e., for which $\s^2_{k+1}-\s^2_k\le\s^2_k-\s^2_{k-1}$, for $k=2,\ldots,n-1$, almost surely has no triple points, and this condition holds for Atlas models and quasi-Atlas models \citep*{IchibaKaratzasShkolnikov:2013,Sarantsev:2015}. Hence, equation~\eqref{2.1} and Lemma~\ref{L2.1} are valid for Atlas models and quasi-Atlas models.

For a first-order model $\{X_1,\ldots,X_n\}$, let us define the processes $\XX_1,\ldots,\XX_n$ by
\[
\XX_i(t)= \log X_i(t)-\frac{1}{n}\sumj\log X_j(t),\quad t\in[0,\infty),
\]
for $i=1,\ldots,n$, along with the corresponding ranked processes $\XX_{(1)}\ge\cdots\ge \XX_{(n)}$, with
\[
\XX_{(k)}(t)= \log X_{(k)}(t)-\frac{1}{n}\sumj\log X_j(t),\quad t\in[0,\infty),
\]
for $k=1,\ldots,n$. Then it follows from Theorems~3.1 and 3.2 of \citet{Khasminskii:1960}, Theorem~4.1 of \citet{Khasminskii:1980}, or Proposition~1 of \citet{IPBKF:2011}, that $(\XX_1,\ldots,\XX_n)$, as a process with values in $\R^n$, has a unique stationary distribution. We define the  {\em gap processes}  by $\log X_{(k)}-\log X_{(k+1)}$, for $k=1,\ldots,n-1$, and the stationary distribution for $(\XX_1,\ldots,\XX_n)$ induces a stationary distribution for each gap process $\log X_{(k)}-\log X_{(k+1)}=\XX_{(k)}-\XX_{(k+1)}$ (see Corollary~2 of \citet{IPBKF:2011}). 

For a first-order model $\{X_1,\ldots,X_n\}$,  let $\xi_k$ represent the gap process $\log X_{(k)}-\log X_{(k+1)}$ in its stationary distribution, for $k=1,\ldots,n-1$.  For an Atlas model or quasi-Atlas model, the $\xi_k$ will be independent and exponentially distributed, so the stationary joint distribution of $(\xi_1,\ldots,\xi_{n-1})$ will be the product of the exponential marginal distributions (this follows from Theorem~9.2 of \citet{Harrison/Williams:1987a} and is a special case of Theorem~2 of \citet{IPBKF:2011}). It is also known that  in this case $\xi_k$ has density function $\a_k e^{-\a_k x}$, for $x\in[0,\infty)$, with rate parameter
\begin{equation}\label{3.00}
\a_k= \frac{2\l_{k,k+1}}{\s^2_{k,k+1}},
\end{equation}
and expectation
\[
\E\big[\xi_k\big]=\frac{1}{\a_k}=\frac{\s^2_{k,k+1}}{2\l_{k,k+1}}
\]
(see   Theorem~2 of \citet{IPBKF:2011}). For $k=1,\ldots,n-1$, if  $f:[0,\infty)\to\R$ is a measurable function with 
\[
\int_0^\infty |f(x)|e^{-\a_k x}dx<\infty,
\]
then the strong law of large numbers
\[
\limT{1}\intT f\big(\log X_{(k)}(t)-\log X_{(k+1)}(t)\big)dt=\E\big[f(\xi_k)\big],\as,
\]
holds (see  Proposition~1  of \citet{IPBKF:2011}, Theorem 3.1 of \citet{Khasminskii:1960}, or Theorem~5.1 on page 121 of \citet{Khasminskii:1980}). It follows from this that
\begin{equation} \label{2.41}
 \limT{1}\intT\big(\log X_{(k)}(t) - \log X_{(k+1)}(t)\big)\,dt = \E\big[\xi_k\big]=\frac{1}{\a_k}=\frac{\s^2_{k,k+1}}{2\l_{k,k+1}},\as,
\end{equation}
and
\begin{equation} \label{2.41a}
 \limT{1}\intT\frac{X_{(k+1)}(t)}{X_{(k)}(t)}\,dt=\E\big[e^{-\xi_{k}}\big]=\frac{\a_k}{\a_{k}+1},\as,
\end{equation}
for $k=1, \ldots, n-1$ (see  Theorem~1 of \citet{IPBKF:2011}). 

For a first-order model $\{X_1,\ldots,X_n\}$, the asymptotic slope of the tangent to the log-log plot of the $X_{(k)}$ versus rank will be
\begin{equation}\label{2.42a}
 \limT{1}\intT \frac{\log X_{(k)}(t) - \log X_{(k+1)}(t)}{\log(k) - \log(k+1)} \, dt
\end{equation}
at rank $k$, so if we define the {\em slope parameters} $s_k$ by
\begin{equation} \label{2.42}
s_k \eqdef  k\limT{1}\intT \big(\log X_{(k)}(t) - \log X_{(k+1)}(t)\big) \, dt,
\end{equation}
for $k = 1, \ldots, n-1$, then
\begin{equation} \label{2.43}
 -s_k\left(1 + \frac{1}{2k}\right) < \limT{1}\intT \frac{\log X_{(k)}(t) - \log X_{(k+1)}(t)}{\log(k) - \log(k+1)} \, dt < -s_k,
\end{equation}
for $k = 1, \ldots, n-1$. Accordingly,  for large enough $k$ the slope parameter   $s_k$ will be approximately equal to minus the slope given in~\eqref{2.42a}. For expositional simplicity, we treat the $s_k$ as if they measured the true log-log slopes between adjacent ranks, but it is important to remember that this equivalence is only as accurate as the range of the inequalities in \eqref{2.43}.

For an Atlas model, it follows from \eqref{2.4}, \eqref{2.41}, and \eqref{2.42} that
\begin{equation}\label{2.44}
s_k = \frac{\s^2}{2g},\as,
\end{equation}
for $k=1,\ldots,n-1$, so the stationary distribution of an Atlas model follows a Pareto distribution, at least within the approximation \eqref{2.43}, and when
\begin{equation*} 
\s^2= 2g,
\end{equation*}
it follows Zipf's law.  For a quasi-Atlas model, we see from  \eqref{2.6}, \eqref{2.7}, and \eqref{2.41} that the slope parameters will be
\begin{equation}\label{2.71}
s_k = \frac{ k\big(\s_k^2+\s^2_{k+1}\big)}{2\l_{k,k+1}}= \frac{ \s_k^2+\s^2_{k+1}}{4g},\as,
\end{equation}
for $k=1,\ldots,n-1$, so the stationary distributions of  quasi-Atlas models are not confined to the class of Pareto distributions. 

It is convenient to consider families of  first-order models that share the same parameters, and for this purpose we define a {\em first-order family}  to be a sequence of constants $\{g_k,\s^2_k\}_{ k\in\N}$, with 
\begin{equation*}\begin{split} 
g_1+\cdots+g_k&<0,\\
\s^2_k&>0,
\end{split}\end{equation*}
for $k\in\N$. A first-order family generates a class of first-order models $\{X_1,\ldots,X_n\}$, each defined as in \eqref{2.5} with the common parameters $g_k$ and $\s^2_k$, for $k\in\N$, with $G_n=-(g_1+\cdots+g_n)$, for $n\in\N$.  An {\em  Atlas family} is a first-order family with $g_k=-g<0$ and $\s^2_k=\s^2>0$, for $k\in\N$. A {\em  quasi-Atlas family} is a first-order family with $g_k=-g<0$ and $\s^2_k=\s_1^2+(k-1)(\s^2_2-\s^2_1)>0$, for $k\in\N$.  

For a first-order family $\{g_k,\s^2_k\}_{k\in\N}$ we use the notation $\E_n$ to denote the expectation with respect to the stationary distribution for the system $\{\log(X_{(1)}/X_{(2)}), \ldots, \log(X_{(n-1)}/X_{(n)})\}$  defined by that family. For Atlas models and quasi-Atlas models it is useful to measure the expected values of the ranked processes $X_{(k)}$ relative to the value of the top process $X_{(1)}$, so we define the {\em ranked weight ratios}
\begin{equation}\label{Xk}
R_k\eqdef \E_n\bigg[\frac{X_{(k)}(t)}{X_{(1)}(t)}\bigg],
\end{equation}
for $k=1,\ldots,n$ and $t\ge0$. Since $\E_n$ assumes the stationary distribution, and since the definition does not depend on weights below the $k$th rank, the ranked weight ratios are independent of both $t$ and $n$. With the system in its stationary distribution, the random variables $\log(X_{(k)}(t)/X_{(k+1)}(t))$ are independent, so
\begin{equation}\label{Xk1}
R_k=\E_n\bigg[\frac{X_{(k)}(t)}{X_{(k-1)}(t)}\bigg]\cdot\E_n\bigg[\frac{X_{(k-1)}(t)}{X_{(k-2)}(t)}\bigg]\cdots\E_n\bigg[\frac{X_{(2)}(t)}{X_{(1)}(t)}\bigg],
\end{equation}
for $2\le k\le n$ and $t\ge0$, where the terms on the right-hand side can be calculated in terms of~\eqref{2.41a}. We can also define, for $n\in\N$,
\begin{equation}\label{Xn}
R_{[n]}\eqdef\E_n\bigg[\frac{X_{[n]}(t)}{X_{(1)}(t)}\bigg]=   R_1+\cdots+  R_n,
\end{equation}
for $t\ge0$.

For an Atlas family or quasi-Atlas family, the parameters $\s^2_{k,k+1}$, $\l_{k,k+1}$, $s_k$, and $R_k$ are defined uniquely for $k\in\N$ by \eqref{2.4}, \eqref{2.49}, \eqref{2.50}, \eqref{2.44}, \eqref{2.71}, and \eqref{Xk}, as the case may be. Let us note that for a quasi-Atlas family the slope parameters $s_k$  and ranked weight ratios $R_k$ do not depend on the number of processes in the model as long as $n>k$, so a quasi-Atlas family defines a unique asymptotic distribution curve.  Accordingly, these families will allow us to derive results about asymptotic distribution curves without repeatedly reciting the characteristics of individual  models. Moreover, we only consider values derived from a first-order family when the models in the family are in their stationary distribution. Hence, for  the Atlas families and quasi-Atlas families we consider, we can calculate the values of the $s_k$ and $R_k$ directly from the parameters $g$, $\s^2_1$, and $\s^2_2$, and we can ignore the models themselves.

\section{Zipfian Atlas models as approximations of empirical systems} \label{zipf}

In this section we first consider how empirical systems of time-dependent data can be approximated by first-order models. In the case that these first-order approximations are in fact Atlas or quasi-Atlas models, we show that it is likely that the empirical systems will follow Zipfian of quasi-Zipfian distributions.

Suppose that  $\{Y_1,\ldots,Y_n\}$, for $n>1$, is an asymptotically stable system of strictly positive continuous semimartingales with rank function $\r_t\in\Sigma_n$  such that $\r_t(i)<\r_t(j)$ if $Y_i(t)>Y_j(t)$ or if $Y_i(t)=Y_j(t)$ and $i<j$. Let $\{Y_{(1)}\ge\cdots\ge Y_{(n)}\}$  be the corresponding rank processes with $Y_{(\r_t(i))}(t)=Y_i(t)$. As in Definition~\ref{D2.2},  for the processes $Y_1\ldots,Y_n$ we can define the parameters
\begin{equation}\label{3.01b}\begin{split}
\bl_{k,k+1}    &\eqdef   \limt{1}\lt{k,k+1}^Y(t) >0,\as,\\ 
\bs^2_{k,k+1}&\eqdef  \limt{1}\bbrac{\log Y_{(k)}-\log Y_{(k+1)}}_t >0,\as,
\end{split}\end{equation}
for $k=1,\ldots,n-1$.

\begin{defn}\label{approx}\citep{F:2002} 
Let $\{Y_1,\ldots,Y_n\}$ be an asymptotically stable system of strictly positive continuous semimartingales with parameters $\bl_{k,k+1}$ and $\bs^2_{k,k+1}$, for $k=1,\ldots,n$, defined by \eqref{3.01b}. Then the {\em first-order approximation} of $\{Y_1,\ldots,Y_n\}$ is the first-order model $\{X_1,\ldots,X_n\}$ with
\begin{equation}\label{3.1}
d\log X_i(t) = g_{r_t(i)}\,dt+G_n\1_{\{r_t(i)=n\}}dt+\s_{r_t(i)}\,dW_i(t),
\end{equation}
for $i=1,\ldots,n$, where $r_t\in\Sigma_n$ is the rank function for the $X_i$, the parameters $g_k$ and $\s_k$ are defined by
\begin{equation}\label{3.2}
\begin{split}
g_k  &= \half \bl_{k-1,k} -  \half \bl_{k,k+1}\text{ for } k=2,\ldots,n-1;\quad g_1=-\half\bl_{1,2}\text{ and } g_n =g_{n-1}\land0;\\
\s_k^2 &= \frac{1}{4}\big(\bs^2_{k-1,k}+\bs^2_{k,k+1}\big)\text{ for } k=2,\ldots,n-1;\\
\s^2_1 &=\s^2_2+\big(\s^2_{2}-\s^2_{3}\big)\1_{\{2\s^2_2>\s^2_3\}} \text{ and } \s^2_n =\s^2_{n-1}+\big(\s^2_{n-1}-\s^2_{n-2}\big)\lor0;
\end{split}
\end{equation}
where $\s_k$ is the positive square root of $\s^2_k$, $G_n=-(g_1+\cdots+g_n)$, and $(W_1,\ldots,W_n)$ is a Brownian motion.
\end{defn}\vspace{5pt}

The parameters $g_1$, $g_n$, $\s^2_1$, and $\s^2_n$ in \eqref{3.2} were chosen to preserve the structure of Atlas models and quasi-Atlas models. For the first-order model \eqref{3.1} with parameters  \eqref{3.2}, equations \eqref{2.6} and \eqref{2.7} imply that
\begin{equation}\label{3.20}
 \l_{k,k+1}=-2\big(g_1+\cdots+g_k\big)= \bl_{k,k+1},\as,
\end{equation}
for $k=1,\ldots,n-1$, and
\begin{equation*} 
\s^2_{k,k+1}=\s^2_k+\s^2_{k+1}
= \frac{1}{4}\big(\bs^2_{k-1,k}+2\bs^2_{k,k+1}+\bs^2_{k+1,k+2}\big),\as,
\end{equation*}
for $k=2,\ldots,n-2$, so the $\s^2_{k,k+1}$ are a smoothed version of the $\bs^2_{k,k+1}$.  Hence, the parameters for a first-order approximation are similar to those of the asymptotically stable system that it approximates. We would also like to have the stable distributions of the two systems $\{\log(X_{(1)}/X_{(2)}),\ldots,\log(X_{(n-1)}/X_{(n)})\}$ and  $\{\log(Y_{(1)}/Y_{(2)}),\ldots,\log(Y_{(n-1)}/Y_{(n)})\}$ be similar, with 
\begin{equation*} 
\limT{1}\intT\big(\log X_{(k)}(t)-\log X_{(k+1)}(t)\big)\,dt \cong \limT{1}\intT\big(\log Y_{(k)}(t)-\log Y_{(k+1)}(t)\big)\,dt,\as,
\end{equation*}
for $k=1,\ldots,n-1$. From \eqref{3.2} and \eqref{3.20} we  see that if the system $\{Y_1,\ldots,Y_n\}$ is a quasi-Atlas model with parameters $\bg_k$ and $\bs^2_k$, then the first-order approximation $\{X_1,\ldots,X_n\}$ will also be a quasi-Atlas model with the same parameters $g_k=\bg_k$ and $\s^2_k=\bs^2_k$, for $k=1,\ldots,n$. In this case it follows from \eqref{2.41} that
\begin{equation*} 
\limT{1}\intT\big(\log X_{(k)}(t)-\log X_{(k+1)}(t)\big)\,dt 
= \limT{1}\intT\big(\log Y_{(k)}(t)-\log Y_{(k+1)}(t)\big)\,dt,\as,
\end{equation*}
for $k=1,\ldots,n-1$, so the stable distributions of the two systems will be the same.

Lemma~\ref{L2.1} shows that the parameters $\bl_{k,k+1}$ can be expressed as
\begin{equation} \label{3.02}
\bl_{k,k+1} 
=\limT{2} \intT\bigg(\frac{dY_{[k]}(t)}{Y_{(k)}(t)}-\sum_{i=1}^n \1_{\{\r_t(i)\le k\}}\frac{dY_i(t)}{Y_{(k)}(t)}\bigg),\as,
\end{equation}
for $k=1,\ldots,n-1$, in which all the terms on the right-hand side of the equation are observable. In a similar fashion we can write
\begin{equation} \label{3.03}
\bs^2_{k,k+1}=\limT{1}\intT  d\bbrac{\log Y_{(k)}-\log Y_{(k+1)}}_t,\as,
\end{equation}
for $k=1,\ldots,n-1$.  These two equations will allow us to define parameters equivalent to $\bl_{k,k+1}$ and $\bs^2_{k,k+1}$  for time-dependent systems of empirical data.

Suppose now that we have a time-dependent system $\{ Z_1(\t), Z_2(\t), \ldots \}$ of positive-valued data observed at times $\t\in\{1, 2, \ldots, T\}$, where $T>1$. Let  
\begin{equation} \label{3.5}
N_\t = \#\{Z_1(\t),Z_2(\t),\ldots\} \quad\text{ and }\quad N=N_{1}\land\cdots\land N_{T}, 
\end{equation}
where $\#$ represents cardinality. Let $\r_\t:\N\to\N$ be the rank function for the system $\{Z_1(\t),Z_2(\t),\ldots\}$ such that $\r_\t$ restricted to the subset $\{1,\ldots,N_\t\}$ is the permutation with $\r_\t(i)<\r_\t(j)$ if $Z_i(\t)>Z_j(\t)$ or if $Z_i(\t)=Z_j(\t)$ and $i<j$, and for $i>N_\t$, $\r_\t(i)=i$. We  define the ranked values $\{Z_{(1)}(\t)\ge Z_{(2)}(\t)\ge\cdots\}$ such that $Z_{(\r_\t(i))}(\t)=Z_i(\t)$ for $i\le N_\t$, and for definiteness we can let $Z_{(k)}(\t)=0$ for $k>N_\t$. With these definitions, we have
\[
Z_{[k]}(\t)=Z_{(1)}(\t)+\cdots+Z_{(k)}(\t),
\]
for $k=1,\ldots,N$ and $\t\in\{1,2,\ldots,T\}$.

We can mimic the time averages \eqref{3.02} and \eqref{3.03} to define the parameters 
\begin{equation} \label{3.6}
\bl_{k,k+1} \eqdef\frac{2}{T-1}\sum_{\t=1}^{T-1}\bigg(\frac{Z_{[k]}(\t+1)-Z_{[k]}(\t)}{Z_{(k)}(\t)}-\sum_{i=1}^N \1_{\{\r_{\t}(i)\le k\}}\frac{Z_i(\t+1)-Z_i(\t)}{Z_{(k)}(\t)} \bigg),
\end{equation}
and
\begin{equation}\label{3.7}
\bs^2_{k,k+1}   \eqdef \frac{1}{T-1}\sum_{\t=1}^{T-1}\Big(\big(\log Z_{(k)}(\t+1)-\log  Z_{(k+1)}(\t+1)\big)
 -\big(\log Z_{(k)}(\t)-\log Z_{(k+1)}(\t)\big)\Big)^2,
\end{equation}
for $k=1,\ldots,N-1$.

\begin{defn}
Suppose that $\{Z_1(\t),Z_2(\t),\ldots\}$, for $T\in\{1,2,\ldots,T\}$, with $T>1$, is a time-dependent system of positive-valued data with $N$, $\bl_{k,k+1}$, and $\bs^2_{k,k+1}$ defined as in \eqref{3.5}, \eqref{3.6}, and \eqref{3.7}. The {\em first-order approximation} of $\{Z_1(\t),Z_2(\t),\ldots\}$ is the first-order family $\{g_k,\s^2_k\}_{ k\in\N}$ with
\begin{equation}\label{3.8}
\begin{split}
g_k  &= \half \bl_{k-1,k} -  \half \bl_{k,k+1}\text{ for } k=2,\ldots,N-1;\quad g_1=-\half\bl_{1,2} \text{ and } g_k =g_{k-1}\land0\text{ for } k\ge N; \\
\s_k^2 &= \frac{1}{4}\big(\bs^2_{k-1,k}+\bs^2_{k,k+1}\big)\text{ for } k=2,\ldots,N-1;\\
\s^2_1 &=\s^2_2+\big(\s^2_{2}-\s^2_{3}\big)\1_{\{2\s^2_2>\s^2_3\}} \text{ and } \s^2_k =\s^2_{k-1}+\big(\s^2_{k-1}-\s^2_{k-2}\big)\lor0\text{ for } k\ge N.
\end{split}
\end{equation}
\end{defn}
\vspace{10pt}

If the first-order model $\{X_1,\ldots,X_{N}\}$ defined by \eqref{3.1} with parameters \eqref{3.8} satisfies
\begin{equation} \label{3.9}
\limT{1}\intT\big(\log X_{(k)}(t)-\log X_{(k+1)}(t)\big)\,dt \cong\frac{1}{T}\sum_{\t=1}^{T}\big(\log Z_{(k)}(\t)-\log Z_{(k+1)}(\t)\big),
\end{equation}
for $k =1,\ldots,N-1$, then we say that the system $\{Z_1(\t),Z_2(\t),\ldots\}$ is {\em rank-based}. If the system $\{Z_1(\t),Z_2(\t),\ldots\}$ is rank-based and the first-order model $\{X_1,\ldots,X_{N}\}$ defined by \eqref{3.1} with parameters \eqref{3.8} is a quasi-Atlas model, then it follows from \eqref{2.41} and \eqref{3.9} that 
\begin{equation} \label{3.10}
\frac{1}{T}\sum_{\t=1}^{T}\big(\log Z_{(k)}(\t)-\log Z_{(k+1)}(\t)\big) \cong \frac{\bs^2_{k,k+1}}{2\bl_{k,k+1}},
\end{equation}
for $k =1,\ldots,N-1$. In this case, the slope parameters for the first-order approximation apply to the distribution curve for the empirical system $\{Z_1(\t),Z_2(\t),\ldots\}$, and this motivates the next two definitions.

\begin{defn} \label{D5.3}
A first-order family  is {\em Zipfian} if its slope parameters $s_k=1$, for $k\in\N$. A time-dependent rank-based system is {\em Zipfian} if its first-order approximation is Zipfian.
\end{defn}

We see that in terms of the parameters $g$ and $\s^2$, an Atlas family is Zipfian if and only if $\s^2=2g$, in which case $\a_k=k$ in \eqref{3.00} and
\begin{equation} \label{3.11}
R_k=\frac{k-1}{k}\cdot\frac{k-2}{k-1}\cdots\frac{1}{2}=\frac{1}{k},
\end{equation}
as in~\eqref{2.41a} and \eqref{Xk1}. Since many empirical distributions are not Zipfian but rather quasi-Zipfian, we need to formalize this concept for first-order families. 

\begin{defn} \label{D5.4}
A first-order family is {\em quasi-Zipfian} if its slope parameters $s_k$  are nondecreasing with $s_1 \leq 1$ and
\begin{equation*} 
\lim_{k\to\infty}s_k \geq 1,
\end{equation*}
where this limit includes divergence to infinity. A time-dependent rank-based system is {\em quasi-Zipfian} if its first-order approximation is quasi-Zipfian.
\end{defn}

For a quasi-Atlas family that is not an Atlas family, we see that in terms of the parameters $g$, $\s^2_1$, and  $\s^2_2$ of \eqref{2.49} and \eqref{2.50}, the family is quasi-Zipfian if and only if $\s^2_1+\s^2_2\le4g$.

By these definitions, a Zipfian system is also quasi-Zipfian. Because the slope parameters $s_k$ are approximately equal to minus the slope of a log-log plot of size versus rank, Definition~\ref{D5.4} implies that a time-dependent rank-based system will be quasi-Zipfian if this log-log plot of its first-order approximation is concave with slope not steeper than $-1$ at the highest ranks and not flatter than $-1$ at the lowest ranks. 

Zipf's law originally referred to the frequency of words in a written language  \citep{Zipf:1935}, with the system $\{Z_1(\t),Z_2(\t),\ldots\}$, where $Z_i(\t)$ represents the number of occurrences of the $i$th word in a language at time~$\t$. To measure the relative frequency of written words in a language it is not possible to observe all the written words in that language. Instead, the words must be {\em sampled,} where a random sample is selected (without replacement), and the frequency versus rank of this random sample is studied. For example, in \citet{wiki-zipf}  10 million words in each of 30 languages were sampled, and the resulting distribution curves created. If the sample is large enough, the distribution of the sampled data should not differ materially from the distribution of the entire data set, at least for the higher ranks.

An advantage that arises from using sampled data is that it is possible to keep the total number of data in the sample constant over time.  The total number of written words that appear in a language is likely to increase over time, and this increase could bias estimates of some parameters. Sampling the data will remove such a trend from the data, since a constant number of words can be sampled at each time.  Accordingly, in all cases we shall assume that global trends have been removed from the data, either by sampling or by some other means of detrending.

Since we have assumed that we have a constant sample size  or that the data have been detrended,  the total count of our sampled data  will remain constant, so
\begin{equation} \label{3.0}
Z_1(\t) + Z_2(\t) + \cdots = \text{ constant},
\end{equation}
for $\t\in\{1,2,\ldots,T\}$, where in the case of the Wikipedia words the constant would be 10 million. 

Suppose we have a time-dependent  system of positive-valued data $\{Z_1(\t),Z_2(\t),\ldots\}$ and we observe the top $n$ ranks, for $1<n<N$, with $N$ from \eqref{3.5}, along with
\[
Z_{[n]}(\t)=Z_{(1)}(\t)+\cdots+Z_{(n)}(\t).
\]
Since the total value of the sampled data in~\eqref{3.0} is constant, for large enough $n$ it is reasonable to expect the relative change of the top $n$ ranks to satisfy
\begin{equation}\label{3.01}
\frac{Z_{[n]}(\t+1)-Z_{[n]}(\t)}{Z_{[n]}(\t)}\cong 0,
\end{equation}
as $n$ becomes large, at least on average over time. This condition is essentially a ``conservation of mass'' criterion for  $\{Z_1(\t),Z_2(\t),\ldots\}$, in which the total ``mass''~\eqref{3.0} of the system remains constant, at least on average over time. It is useful to normalize the values $Z_{(k)}(\t)$ and $Z_{[n]}(\t)$ by measuring them relative to the largest value $Z_{(1)}(\t)$, in which case~\eqref{3.01} becomes  
\begin{equation*} 
\frac{1}{\big(Z_{[n]}(\t)/Z_{(1)}(\t)\big)}\frac{Z_{[n]}(\t+1)-Z_{[n]}(\t)}{Z_{(1)}(\t)}\cong 0,
\end{equation*}
as $n$ becomes large, at least on average over time. For the first-order family $\{g_k,\s^2_k\}_{k\in\N}$, this expression allows us to use the ranked weight ratios $R_k$ and $R_{[n]}$  of~\eqref{Xk} and \eqref{Xn}, and motivates the following definition.

\vspace{5pt}
\begin{defn} \label{D5.1}
The first-order family $\{g_k,\s^2_k\}_{k\in\N}$ is {\em conservative} if for $T>0$,
\begin{equation*} 
\lim_{n\to\infty}\frac{1}{  R_{[n]}} \E_n\bigg[\frac{1}{T}\intT\frac{dX_{[n]}(t)}{X_{(1)}(t)}\bigg]= 0.
\end{equation*}
\end{defn}
\vspace{5pt}

For the system $\{Z_1(\t),Z_2(\t),\ldots\}$ and for $n<N$, the replacement of processes in the top $n$ ranks  by processes in the lower ranks over the time interval $[\t,\t+1]$ is measured by
\[
Z_{[n]}(\t+1)-\sum_{i=1}^N\1_{\{\r_{\t}(i)\le n\}}Z_i(\t+1),
\]
or
\[
\big(Z_{[n]}(\t+1)-Z_{[n]}(\t)\big)-\bigg(\sum_{i=1}^N\1_{\{\r_{\t}(i)\le n\}}\big(Z_i(\t+1)-Z_i(\t)\big)\bigg).
\]
While some replacement from lower ranks is necessary, it seems reasonable to expect that the system will be ``complete'' in the sense that on average the relative proportion of the mass that is replaced becomes arbitrarily small for large enough $n$, i.e., that
\begin{equation*} 
\frac{1}{Z_{[n]}(\t)}\Big(Z_{[n]}(\t+1)-Z_{[n]}(\t)-\sum_{i=1}^N\1_{\{\r_{\t}(i)\le n\}}\big(Z_i(\t+1)-Z_i(\t)\big)\Big)\cong 0,
\end{equation*}
for large enough $n$. As in Definition~\ref{D5.1}, in terms of the first-order approximation of  $\{Z_1(\t),Z_2(\t),\ldots\}$, this becomes
\begin{equation} \label{5.10}
\frac{1}{  R_{[n]}} \E_n\bigg[\frac{1}{T}\intT\frac{dX_{[n]}(t)}{X_{(1)}(t)}-\frac{1}{T}\intT\bigg(\sum_{i=1}^N\1_{\{r_{t}(i)\le n\}}\frac{dX_i(t)}{X_{(1)}(t)}\bigg)\bigg]\cong 0,
\end{equation}
for $T>0$ and large enough $n$, where $N>n$ and $\{X_1,\ldots,X_N\}$ is a first-order model defined by $\{g_k,\s^2_k\}_{k\in\N}$. By \eqref{2.2}, this is equivalent to
\[
\frac{1}{  R_{[n]}} \E_n\bigg[\frac{1}{T}\intT\frac{X_{(n)}(t)}{2X_{(1)}(t)}d\lt{n,n+1}^X(t)\bigg]\cong 0,
\]
for $T>0$ and large enough $n$. Since
\begin{equation*} 
\lim_{T\to\infty}\frac{1}{T}\intT d\lt{n,n+1}^X(t)=\l_{n,n+1}=-2\big(g_1+\cdots+g_n\big),\as,
\end{equation*}
condition \eqref{5.10} corresponds to
\[
\frac{1}{  R_{[n]}} \E_n\bigg[\frac{1}{T}\intT-\big(g_1+\cdots+g_n\big)\frac{X_{(n)}(t)}{X_{(1)}(t)}dt\bigg]\cong 0,
\]
for $T>0$ and large enough $n$. Since $\E_n$ assumes the stationary distribution, this is equivalent to
\[
-\big(g_1+\cdots+g_n\big)\frac{   R_n}{  R_{[n]}}\cong 0,
\]
for large enough $n$, and with $G_n=-\big(g_1+\cdots+g_n\big)$, we have the following definition.

\begin{defn} \label{D5.2}
The first-order family $\{g_k,\s^2_k\}_{ k\in\N}$ is {\em complete} if 
\begin{equation*} 
\lim_{n\to\infty} \frac{G_n   R_n}{  R_{[n]}}= 0.
\end{equation*}
\end{defn}\vspace{5pt}

For an Atlas family or quasi-Atlas family $G_n=ng$, so for these families completeness is equivalent to
\begin{equation*} 
\lim_{n\to\infty} \frac{n  R_n}{R_{[n]}}= 0.
\end{equation*}
The following two propositions show that conservation and completeness are the basis for the Zipfian nature of the distributions of many systems of time-dependent rank-based data.

\begin{prop} \label{T1}
An Atlas family is Zipfian if and only if it is conservative and complete.
\end{prop}

This proposition has  a natural counterpart for quasi-Atlas families.

\begin{prop} \label{T2}
If a quasi-Atlas family is conservative and complete with
\begin{equation} \label{5.5}
\lim_{n\to\infty} R_{[n]} \ge 2,
\end{equation}
then it is quasi-Zipfian. 
\end{prop}

These two propositions seem remarkably simple. Many empirical systems can be at least roughly approximated by quasi-Atlas models, and conservation and completeness are properties that are  almost universal in large time-dependent rank-based systems of empirical data. If these conditions are satisfied, then these two propositions show that Zipf's law, or at least its quasi-Zipfian counterpart, will pertain. Perhaps it is this simplicity that leads to the universality of Zipf's law for these systems. 


\section{Examples and discussion} \label{disc}

\begin{figure}[H]
\begin{center}
\vspace{-55pt}
\hspace{-20pt}\scalebox{.8}{{\includegraphics{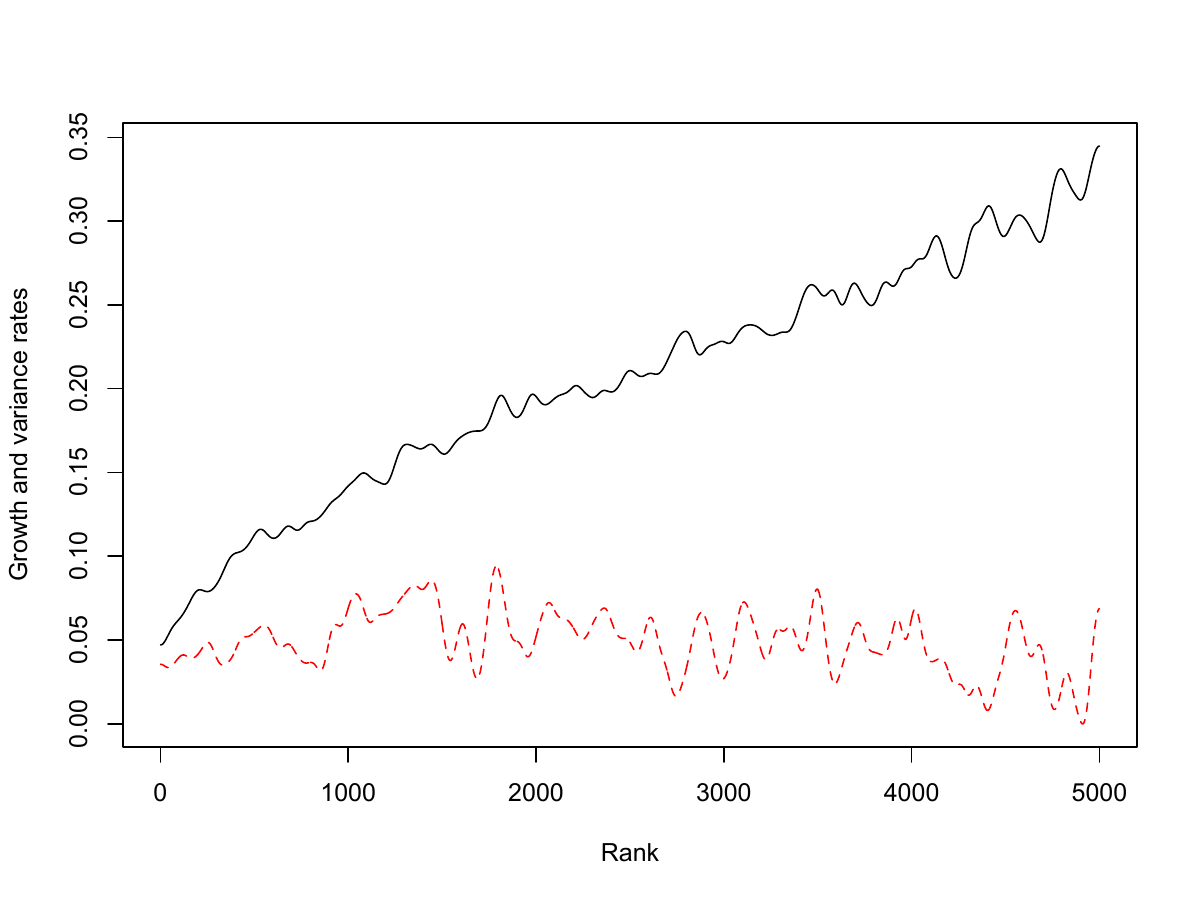}}}
\end{center}
\vspace{-24pt} \caption{U.S.\ capital distribution first-order parameters (smoothed):  $\s^2_k$ (black), $-g_k$ (red, broken).}\label{f1}
\end{figure}

\begin{figure}[H]
\begin{center}
\vspace{-35pt}
\hspace{-20pt}\scalebox{.8}{{\includegraphics{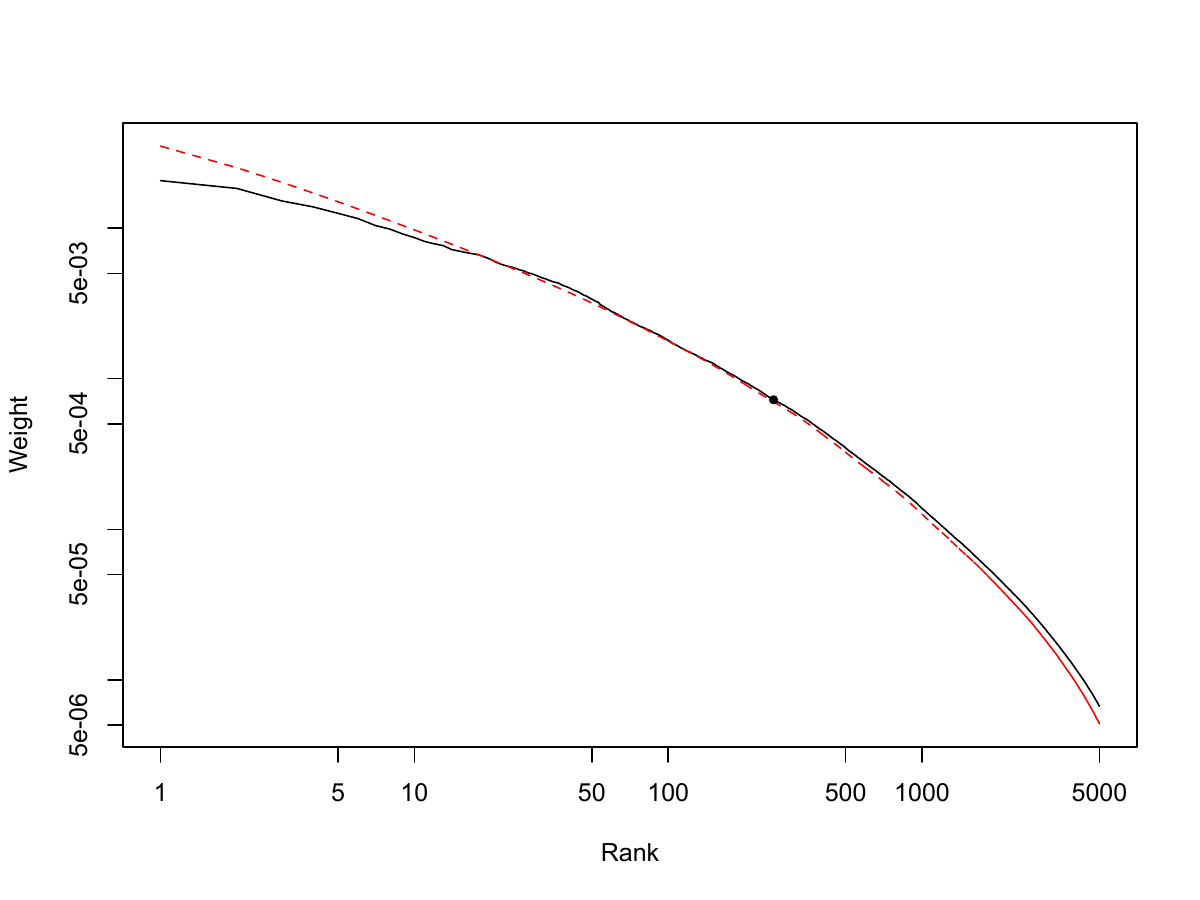}}}
\end{center}
\vspace{-24pt} \caption{ U.S.\ capital distribution, 1990--1999 (black). First-order approximation (red, broken).}\label{f2}
\end{figure}

Empirical time-dependent systems often behave like quasi-Atlas families, and in the Example~\ref{ex1} below we consider one such system, the capitalizations of U.S.\ companies (see Figures~\ref{f1} and \ref{f2}). The condition that the variance rates increase with rank seems natural --- even in the original observation of \citet{Brown:1827} it would seem likely that the water molecules would have buffeted the smaller particles more vigorously than the larger ones. Below the top few ranks, the members of empirical time-dependent systems constantly drift among nearby ranks, and this could result in linearity of the $\s^2_k$, at least throughout the middle ranks. Whether the $g_k=-g$ for all $k$ may be more problematic, but this appears to hold at least in Example~\ref{ex1}, where we analyze actual data. Since we are usually observing the top part of a larger distribution, there is ``leakage'' out of the system, characterized by the last term in \eqref{2.2}, so the constant $-g$ may represent the universal draw toward extinction in time-dependent rank-based systems.

\begin{exa}\label{ex1}  {\em Market capitalization of companies.}
\vspace{5pt}

The market capitalization of U.S.\ companies was studied  as early as \citet{Simon/Bonini:1958}, and here we follow the methodology of \citet{F:2002}. The capitalization of a company is defined as the price of the company's stock multiplied by the number of shares outstanding. Ample data are available for stock prices, and this allows us to estimate the first-order parameters we introduced in the previous sections.

Figure~\ref{f1} shows the smoothed first-order parameters $\s^2_k$ and $-g_k$ for the U.S.\ capital distribution for the 10 year period from January 1990 to December 1999. The capitalization data we used were from the monthly stock database of the Center for Research in Securities Prices at the University of Chicago. The market we consider consists of the stocks traded on the New York Stock Exchange, the American Stock Exchange, and the NASDAQ Stock Market, after the removal of all Real Estate Investment Trusts, all closed-end funds, and those American Depositary Receipts not included in the S\&P 500 Index. The parameters in Figure~\ref{f1} correspond to the 5000 stocks with the highest capitalizations each month. The first-order parameters $g_k$ and  $\s^2_k$ were calculated as in \eqref{3.8} from the parameters $\bl_{k,k+1}$ and $\bs^2_{k,k+1}$ of \eqref{3.6} and \eqref{3.7}, and then smoothed by convolution  with a Gaussian kernel with $\pm 3.16$ standard deviations spanning 100 months on the horizontal axis, with reflection at the ends of the data.

We see in Figure~\ref{f1} that the values of the parameters $-g_k$ are relatively constant compared to the parameters $\s^2_k$, which increase almost linearly with rank. The near-constant $-g_k$ and near-linearly increasing $\s^2_k$ suggest that the first-order approximation can be represented by a quasi-Atlas family. In Figure~\ref{f2}, the distribution curve for the capitalizations is represented by the black curve, which represents the average of the year-end capital distributions for the ten years spanned by the data. The broken  red curve is the first-order approximation of the distribution following \eqref{3.10}. The  two curves are quite close, and this indicates that the time-dependent system of company capitalizations seems to be  rank-based. The black dot on the curve between ranks 100 and 500 is the point at which the log-log slope of the tangent to the curve is $-1$, so this is a quasi-Zipfian distribution, consistent with Proposition~\ref{T2}. Note that if we had considered only the top 100 companies, the completeness condition, Definition~\ref{D5.2}, would have failed, as we would expect for an incomplete distribution.
\end{exa}

\begin{exa}  {\em Frequency of written words.}
\vspace{5pt}

Word frequency is  the origin of Zipf's law \citep{Zipf:1935}, but testing our methodology with word-frequency could be difficult.  Ideally, we would like to construct a first-order approximation for the data and compare the first-order distribution to that of the original data. However, the parameters $\bl_{k,k+1}$ and $\bs^2_{k,k+1}$ for the top-ranked words in a language are likely to be difficult to estimate over any reasonable time frame, since the top-ranked words probably seldom change ranks. Nevertheless, while the top ranks may require centuries of data for accurate estimates, the lower ranks could  be amenable to analysis similar to that which we carried out for company capitalizations.  Moreover, it might be possible to combine, for example, all the Indo-European languages and generate accurate estimates of the $\bl_{k,k+1}$ and $\bs^2_{k,k+1}$ even for the top ranks of the combined data.

We can see from the remarkable chart in \citet{wiki-zipf} that the log-log plots for 30 different languages are (almost) straight. Actually, these plots seem to be slightly concave, or quasi-Zipfian in nature. It is possible that this slight curvature is due to sampling error at the lower ranks, which would raise the variances and steepen the slope, but this would have to be determined by studying the actual data.
\end{exa}

\begin{exa}  {\em Random growth processes.}
\vspace{5pt}

Economists have traditionally used random growth processes to model time-dependent systems with quasi-Zipfian distributions. For example, these processes were used by \citet{Gabaix:1999} to model the distribution of city populations and by \citet{Piketty:2017} to construct a piecewise approximation to the distribution curves for the income and wealth of U.S.\ households. A {\em random growth process} is  an \ito\ process of the form
\begin{equation}\label{rg0}
\frac{dX(t)}{X(t)}=\mu(X(t))\,dt+\s(X(t))\,dW(t),
\end{equation}
where $W$ is Brownian motion and $\m$ and $\s$ are well-behaved real-valued functions. We can convert this into logarithmic form by \ito's rule, in which case
\begin{equation}\label{rg1}
d\log X(t) = \bigg(\m(X(t))-\frac{\s^2(X(t))}{2}\bigg)dt+\s(X(t))\,dW(t), \as
\end{equation}
We shall assume that this equation has at least a weak solution with $X(t)>0$, a.s., and that the solution has a stationary distribution.

Let us construct $n$ i.i.d.\ copies $X_1,\ldots,X_n$ of $X$, all defined by \eqref{rg0} or, equivalently, by \eqref{rg1}, and assume that the $X_i$ are all in their common stationary distribution. Let us assume that the $\log X_i$ accumulate no local time at triple points, so we can define the rank processes and \eqref{2.1} and \eqref{2.2} will be valid. If the system is asymptotically stable we can calculate the corresponding rank-based growth rates $g_k$, but if we know the stationary distribution of the original process \eqref{rg0}, then there is a simpler way to proceed.

If we know the common stationary distribution of the $X_i$, then we can calculate expectations under this stationary distribution and let
\begin{equation}\label{rg2}
g_k= \E\bigg[\m(X_{(k)}(t))-\frac{\s^2(X_{(k)}(t))}{2}\bigg]\quad\text{ and }
\quad \s^2_k= \E\big[\s^2(X_{(k)}(t))\big],
\end{equation}
for $k=1,\ldots,n$. Under appropriate regularity conditions on the $\m$ and $\s$, the expectations here will be equal to the asymptotic time averages of the functions. Since the $X_i$ are in their stationary distribution, the geometric mean $\big(X_1X_2 \ldots X_n\big)^{1/n}=\big(X_{(1)}X_{(2)}\ldots X_{(n)}\big)^{1/n}$ will also be in its stationary distribution, so
\begin{equation*}
\big(g_1+\cdots+g_n\big)t=\E\big[\log \big(X_{(1)}(t)\cdots\ X_{(n)}(t)\big)-\log \big(X_{(1)}(0)\cdots\ X_{(n)}(0)\big)\big]=0.
\end{equation*}
Hence,
\begin{equation}\label{rg2.1}
g_1+\cdots+g_n=0,\quad\text{ with }\quad g_1+\cdots+g_k<0,\text{ for }  k<n,
\end{equation}
so the $g_k$ and $\s^2_k$  define the first-order model
\begin{equation}\label{rg3}
d\log Y_i(t)= g_{r_t(i)}dt+\s_{r_t(i)}dW_i(t),
\end{equation}
for $i=1,\ldots,n$, where $W_1,\ldots,W_n$ is $n$-dimensional Brownian motion. In this case,  $G_n=0$.

If the functions $\m$ and $\s$ in \eqref{rg0} are smooth enough, then the system is likely to be rank-based, with  the stationary distribution of the first-order model \eqref{rg3}  close to that of the original system \eqref{rg0}. More conditions are required for this stationary distribution to be quasi-Zipfian, and to achieve a true Zipfian distribution, a lower reflecting barrier or other equivalent device must be included in the model \citep{Gabaix:2009}.
\end{exa}

\begin{exa}  {\em Population of cities.}
\vspace{5pt}

The distribution of city populations is a prominent example of Zipf's law in social science. However, as the comprehensive cross-country investigation of \citet{Soo:2005} shows, city size distributions in most countries are not Zipfian but rather quasi-Zipfian. \citet{Gabaix:1999} hypothesized that the  quasi-Zipfian distribution of U.S.\ city size was caused by higher population variances at the lower ranks, consistent with Proposition~\ref{T2}. Which of the deviations from  Zipf's law uncovered by \citet{Soo:2005} are due to population variances that increase with decreasing city size remains an open question.

There is another phenomenon that occurs with city size distributions. Suppose that rather than studying a large country like the U.S.\, we consider instead the populations of the cities in New York State. According to the 2010 U.S.\ census, the largest city, New York City, had a population of 8,175,133, while the second largest, Buffalo, had only 261,310, so this distribution is non-Zipfian. The corresponding population of New York State was 19,378,102, so hypothesis \eqref{5.5} of Proposition~\ref{T2} is satisfied, but nevertheless the proposition fails. This calls for an explanation, and we conjecture that while the population of the cities of New York State comprise a time-dependent system, this system is not rank-based. The population of New York City is not determined merely by its rank among New York State cities, but is highly city-specific in nature. Hence, we cannot expect the stationary distribution for the gap process between New York City and second-ranked Buffalo to be exponential, and we cannot expect the distribution of the system to be quasi-Zipfian.
\end{exa}

\begin{exa}  {\em Assets of banks.}
\vspace{5pt}

\citet{Fernholz/Koch:2016a} show that the distribution of assets held by U.S.\ bank holding companies, commercial banks, and savings and loan associations are all quasi-Zipfian. This is true despite the fact that these distributions have undergone significant changes over the past few decades. However, as \citet{Fernholz/Koch:2017} show, the first-order approximations of these time-dependent rank-based systems generally do not satisfy the hypotheses of Proposition~\ref{T2}, since the parameters $\bs^2_{k,k+1}$ are, in most cases, lower for higher values of $k$. Nonetheless, the parameters $\bl_{k,k+1}$ vary with $k$ in such a way as to generate quasi-Zipfian distributions.
\end{exa}

\begin{exa}  {\em Employees of firms.}
\vspace{5pt}

\citet{Axtell:2001} shows that the distribution of employees of U.S.\ firms is close to Zipfian, with only slight concavity.  A number of empirical analyses have shown that for all but the tiniest firms, employment growth rates of U.S.\ firms do not vary with firm size \citep{Neumark/Wall/Zhang:2011}. This observation together with the slight concavity demonstrated by \citet{Axtell:2001} suggests that the first-order approximation of U.S.\ firm employees might be a quasi-Atlas family, which would explain its quasi-Zipfian nature.
\end{exa}

\section{Conclusion} \label{concl}
We have shown that the stationary distribution of an Atlas family will follow Zipf's law if and only if the family is conservative and complete.  We have also shown that a quasi-Atlas family will have a quasi-Zipfian stationary distribution if the family is conservative and complete, provided that the largest member does not represent more than one half of the total weight of the family. Since conservation and completeness are natural conditions for systems of time-dependent rank-based empirical data, and since many such systems can be approximated by Atlas families or quasi-Atlas families, our results offer an explanation for the universality of Zipf's law for these systems.

\vspace{20pt}
\noindent
{\bf Acknowledgments.} We thank Xavier Gabaix, Ioannis Karatzas, members of the Intech SPT seminar, and participants of the 2017 Thera Stochastics Conference for their invaluable comments and suggestions regarding this research. We are also grateful to an anonymous referee for pointing out a significant error in the original manuscript that led to a major revision of the paper.

\appendix
\section{Proofs}

\noindent{\bf Proof of Lemma \ref{L2.1}.}
Suppose that  the rank processes $X_{(k)}$ satisfy \eqref{2.1}, so we have
\begin{equation*}
d\log X_{(k)}(t) = \sum_{i=1}^{n} \1_{\{r_t(i)=k\}}d\log X_i(t)
+\half d\lt{k,k+1}^X(t)-\half d\lt{k-1,k}^X(t),\as,
\end{equation*}
for $k=1,\ldots,n$. By \ito's rule this is equivalent to
\begin{align*}
\frac{dX_{(k)}(t)}{X_{(k)}(t)}&= \sum_{i=1}^{n} \1_{\{r_t(i)=k\}}\frac{dX_i(t)}{X_{i}(t)}
+\half d\lt{k,k+1}^X(t)-\half d\lt{k-1,k}^X(t)\\
 &= \sum_{i=1}^{n} \1_{\{r_t(i)=k\}}\frac{dX_i(t)}{X_{(k)}(t)}
+\half d\lt{k,k+1}^X(t)-\half d\lt{k-1,k}^X(t),\as,
\end{align*}
for $k=1,\ldots,n$. From this we have
\begin{align*}
dX_{(k)}(t) &= \sum_{i=1}^{n} \1_{\{r_t(i)=k\}}dX_i(t)
+\half X_{(k)}(t)d\lt{k,k+1}^X(t)-\half X_{(k)}(t)d\lt{k-1,k}^X(t)\\
&= \sum_{i=1}^{n} \1_{\{r_t(i)=k\}}dX_i(t)
+\half X_{(k)}(t)d\lt{k,k+1}^X(t)-\half X_{(k-1)}(t)d\lt{k-1,k}^X(t),\as,
\end{align*}
for $k=1,\ldots,n$, since the support of $d\lt{k-1,k}^X$ is contained in the set $\big\{t:\log X_{(k-1)}(t)=\log X_{(k)}(t)\big\}$. Now we can add up $dX_{(1)}(t)+\cdots+dX_{(k)}(t)=dX_{[k]}(t)$ and we have
\begin{equation*}
dX_{[k]}(t)= \sum_{i=1}^{n} \1_{\{r_t(i)\le k\}}dX_i(t)+\half X_{(k)}(t)d\lt{k,k+1}^X(t),\as,
\end{equation*}
for $k=1,\ldots,n$, and \eqref{2.2} follows. \qed

\vspace{10pt}
\noindent{\bf Proof of Proposition \ref{T1}.}
For an Atlas model $\{X_1,\ldots,X_n\}$ with parameters $g>0$ and $\s>0$, \ito's rule implies that
\begin{equation*}
dX_i(t) = \bigg( \frac{\s^2}{2}-g+ng\1_{\{r_t(i)=n\}}\bigg)X_i(t)\,dt+\s X_i(t)\,dW_i(t),\as,
\end{equation*}
for $i = 1, \ldots, n$. Hence,
\begin{equation*}
dX_{[n]}(t)=\bigg(\frac{\s^2}{2}-g\bigg)X_{[n]}(t)\,dt+ X_{[n]}(t) \, dM(t) +ngX_{(n)}(t)\,dt, \as,
\end{equation*}
where $M$ is a local martingale incorporating all of the terms $\s \,dW_i(t)$.  From this we have
\begin{equation*}
\frac{dX_{[n]}(t)}{X_{(1)}(t)} = \bigg(\frac{\s^2}{2}-g\bigg)\frac{X_{[n]}(t)}{X_{(1)}(t)}\,dt + \frac{X_{[n]}(t)}{X_{(1)}(t)} \,dM(t) + \frac{ngX_{(n)}(t)}{X_{(1)}(t)}\,dt, \as,
\end{equation*}
so, for $T>0$,
\begin{equation*}
\E_n\bigg[\frac{1}{T}\intT\frac{dX_{[n]}(t)}{X_{(1)}(t)}\bigg]= \bigg(\frac{\s^2}{2}-g\bigg)  R_{[n]}+ng  R_n,
\end{equation*}
or,
\begin{equation}\label{AA1}
\frac{1}{  R_{[n]}} \E_n\bigg[\frac{1}{T}\intT\frac{dX_{[n]}(t)}{X_{(1)}(t)}\bigg]= \frac{\s^2}{2}-g+\frac{ng  R_n}{  R_{[n]}}.
\end{equation}
If an Atlas family is conservative and complete, then as $n$ tends to infinity the first and last terms of \eqref{AA1} converge to zero,  so $\s^2/2g = 1$ and the family will be Zipfian. 

If the Atlas family is  Zipfian then $\s^2/2g=1$, in which case \eqref{3.11} holds, so
\[
R_k=\frac{1}{k},
\]
and
\[
  R_{[n]}=\sum_{k=1}^n \frac{1}{k}=O(\log n).
\]
It follows that
\[
\frac{ng  R_n}{  R_{[n]}}=\frac{g}{O(\log n)},
\]
so the family is complete, and with $\s^2/2=g$ the right hand side of  \eqref{AA1} converges to zero as $n$ tends to infinity.  Hence, the left-hand side must also converge to zero, so the family is conservative. \qed

\vspace{10pt}
\noindent{\bf Proof of Proposition \ref{T2}.}
Let $\{X_1,\ldots,X_n\}$ be a quasi-Atlas model  with parameters  $g,\s^2_1>0$ and $\s^2_2\ge\s^2_1$, such that $g_k=-g$ and $\s^2_k=\s^2_1+(k-1)(\s^2_2-\s^2_1)$, for $k=1,\ldots,n$.   \ito's rule implies that
\begin{equation*}
dX_i(t) = \bigg( \frac{\s^2_{r_t(i)}}{2} - g+ng\1_{\{r_t(i)=n\}}\bigg)X_i(t)\,dt+\s_{r_t(i)}  X_i(t)\,dW_i(t),\as,
\end{equation*}
for $i = 1, \ldots, n$, so
\begin{equation*}
dX_{[n]}(t) = \sumk X_{(k)}(t)\bigg(\frac{\s^2_k}{2}-g\bigg)dt + dM(t) +ngX_{(n)}(t)\,dt, \as,
\end{equation*}
where $M$ is a local martingale incorporating all of the terms $\s_{r_t(i)}X_i(t) \, dW_i(t)$. As with \eqref{AA1} above, for $T>0$, 
\begin{equation*} 
\frac{1}{  R_{[n]}} E_n\bigg[\frac{1}{T}\intT\frac{dX_{[n]}(t)}{X_{(1)}(t)}\bigg]= \frac{1}{  R_{[n]}} \sumk  R_k\bigg(\frac{\s^2_k}{2}-g\bigg)+\frac{ng  R_n}{  R_{[n]}}.
\end{equation*}
Since the family is conservative and  complete, the first and last terms of this equation converge to zero as $n$ tends to infinity, so 
\begin{equation} \label{D5,2ProofEq}
 \lim_{n\to\infty}\bigg(\frac{1}{  R_{[n]}} \sumk  R_k\frac{\s^2_k}{2g}\bigg) =1.
\end{equation}

Let us now show that \eqref{5.5} implies that $s_1\le1$. Since  $0<\s^2_1\le\cdots\le\s^2_n$, equation  \eqref{D5,2ProofEq} implies that 
{\allowdisplaybreaks
\begin{align*}
 1& \ge \lim_{n\to\infty}\frac{1}{  R_{[n]}}\frac{\s_1^2}{2g}+ \lim_{n\to\infty} \bigg(\frac{1}{  R_{[n]}} \sum_{k=2}^n  R_k\frac{\s^2_2}{2g}\bigg)\\
 &=\lim_{n\to\infty}\frac{1}{  R_{[n]}}\frac{\s_1^2}{2g} +\bigg(1- \lim_{n\to\infty} \frac{1}{  R_{[n]}}\bigg)\frac{\s_2^2}{2g}\\
& \ge \half \frac{\s_1^2}{2g}+\half \frac{\s_2^2}{2g}=s_1,
\end{align*}
}
where the last inequality follows from \eqref{5.5}.

We must now show that either $\lim_{k\to\infty}s_k\ge 1$ or the $s_k$ diverge to infinity. Since the $\s^2_k$ are nondecreasing, as $k$ tends to infinity they must either converge to a finite value $\s^2>0$ or diverge to infinity. We see from \eqref{2.71} that if the $\s^2_k$ diverge to infinity, the same will be true for the $s_k$.  If $\lim_{k\to\infty}\s^2_k=\s^2$ then $\lim_{k\to\infty}s_k=\s^2/2g$, and since the $\s^2_k$ are nondecreasing, 
\[
1=   \lim_{n\to\infty}\bigg(\frac{1}{  R_{[n]}} \sumk  R_k\frac{\s^2_k}{2g} \bigg)\le \frac{\s^2}{2g}.
\]
It follows that $\lim_{k\to\infty}s_k\ge 1$. \qed


\bibliographystyle{chicago}
\bibliography{math3}

\pagebreak

\end{document}